
\input harvmac
\line{\hfill \hfill TIT/HEP-234}
\line{\hfill BNL-49435}
\line{\hfill YNU-HEPTh-93-103}
\line{\hfill hep-ph/9309287}
\line{\hfill September, 1993}
\bigskip
\bigskip
\bigskip
\bigskip
\centerline{\bf MUONIUM HYPERFINE STRUCTURE and THE DECAY
$\mu ^+\to e^+ +\overline \nu _e +\nu _{\mu}$}
\centerline{\bf IN MODELS WITH DILEPTON GAUGE BOSONS}
\bigskip
\bigskip
\bigskip
\medskip
\centerline{H.~Fujii$^1$, Y.~Mimura$^2$
\footnote{$^\sharp$}{e-mail address: mim@phys.titech.ac.jp},
K.~Sasaki$^1$ \footnote{$^\dagger$}
{e-mail address: a121004@c1.ed.ynu.ac.jp}, and
T.~Sasaki$^2$ \footnote{$^\flat$}
{e-mail address: tsasaki@phys.titech.ac.jp}}
\medskip
\bigskip
\centerline{$^1$\sl Dept. of Physics, Faculty of Education}
\centerline{\sl Yokohama National University, Yokohama 240, Japan}
\bigskip
\centerline{$^2$\sl Dept. of Physics, Tokyo Institute of Technology}
\centerline{\sl Oh-okayama, Meguro, Tokyo 152, Japan}

\bigskip
\bigskip
\bigskip
\medskip

\vskip 0.5in
\centerline{\bf Abstract}
\smallskip
\noindent

We examine the muonium ($\mu ^+e^-$)-antimuonium ($\mu ^-e^+$) system
in the models which
accomodate the dilepton gauge bosons, and study their contributions
to the ground state hyperfine splitting in ``muonium''. We also consider
the exotic muon decay $\mu ^+\to e^+ +\overline \nu _e +\nu _{\mu}$
mediated by the dilepton gauge boson, and obtain a lower bound
$(M_{X^{\pm }}/g_{3l})>550 \rm GeV$ at 90\% confidence level for
the singly-charged dilepton mass using the unitarity relation of
the Kobayashi-Maskawa matrix for the 3-family case.


\baselineskip=20pt plus 2pt minus 2pt
\vfill\eject

There have been proposed up to now
a variety of unified models which extend the Standard Model of strong and
electroweak interactions.
Among them there is a class of theories
\ref\Adler{S. L. Adler, Phys. Lett. B{\bf 225}, 143 (1989).}%
\nref\Frama{P. H. Frampton and B.-H. Lee, Phys. Rev. Lett. {\bf 64}, 619
(1990).}%
\nref\Framc{P. H. Frampton and T. W. Kephart, Phys. Rev. D{\bf 42},
3892 (1990).}%
\nref\Pisa{F. Pisano and V. Pleitez, Phys. Rev. D{\bf 46}, 410 (1992).}%
\nref\Framd{P. H. Frampton , Phys. Rev. Lett. {\bf 69}, 2889 (1992).}%
--\ref\Ng{D. Ng, Preprint No. TRI-PP-92-125 (1992).}
in which there appear $SU(2)_L$-doublet gauge bosons
$(X^{\mp}, X^{\mp \mp})$ carrying
lepton number $L=\pm 2$.
Hereafter, we refer these gauge bosons as dileptons.
In these models each family of leptons $(l^+, \nu _l, l^- )_L$
transforms as a triplet under the gauge group $SU(3)$ and
the total lepton number
defined as $L=L_e+L_\mu +L_\tau $ is conserved, while the separate lepton
number
for each family is not. The gauge group $SU(3)$ will be, for example,
a $SU(3)_l$ in the $SU(15)$ grand unification theory (GUT) model \Frama\ or
a $SU(3)_L$ in the $SU(3)_L \times U(1)_X$ model \Framd.

The dilepton masses are generated when the SU(3) gauge symmetry
is spontaneously broken at the scale which could be as low as
250-2000 GeV\Frama
\Framd
\ref\Pal{P. B. Pal, Phys. Rev. D{\bf 43}, 236 (1991).},
and thus they may possibly be
in the range accessible to the future accelerator experiments.
The signatures for the existence of dileptons have been studied
for the future experiments concerning
$e^+ e^-$, $e^- e^-$ and $e^- p$ collisions
\ref\Framb{P. H. Frampton and D. Ng, Phys. Rev. D{\bf 45}, 4240 (1992);
 P. H. Frampton, Mod. Phys. Lett. A{\bf 7}, 2017 (1992).}%
\ref\Rizzo{T. G. Rizzo, Phys. Rev. D{\bf 46}, 910 (1992) }%
\ref\Agr{J. Agrawal, P. H. Frampton and D. Ng, Nucl. Phys. B{\bf 386},
267 (1992).}.
Also, phenomenology on dileptons has been analyzed in such processes as
the low-energy weak neutral current experiments,
Bhabha $e^+ e^-$ scattering, muon decays \Framb
\ref\Carlson{E. D. Carlson and P. H. Frampton, Phys. Lett. B{\bf 283},
123 (1992).}
,
and the low energy muon-related processes
\ref\Fujii{H. Fujii, S. Nakamura, and K. Sasaki, Phys. Lett. B{\bf 299}, 342
(1993).}.
Furthermore, the dilepton contributions to so-called oblique corrections have
been examined recently
\ref\Sasaki{K. Sasaki, Phys. Lett. B{\bf 308}, 297 (1993).}%
\ref\Liu{J. T. Liu and D. Ng, Preprint No. IFP-460-UNC:TRI-PP-93-11 (1993).}.
The most stringent lower mass-bounds at present are
$(M_{X^{\pm \pm}}/g_{3l})>340\rm GeV$ (95\% C.L.) for the doubly-charged
dileptons \Framb\ and $(M_{X^{\pm }}/g_{3l})>640\rm GeV$ (90\% C.L.)
for the singly-charged ones
\Carlson, and here $g_{3l}$ is the gauge coupling of dileptons to leptons.

In Ref.\Fujii, two of the present authors (H.F.and K.S)
and Nakamura have analysed the contributions of dileptons to the low energy
muon-related processes such as the muonium-antimuonium
conversion, the second-order corrections to the muon anomalous
magnetic moment and the exotic muon decay
$\mu ^+\to e^+ +\overline \nu _e +\nu _{\mu}$, and evaluated the lower
bounds on the dilepton mass.
In this paper, we examine again the muonium-antimuonium system
in the models which accomodate the dilepton gauge bosons, and study the
dilepton contributions to the ground state hyperfine splitting in
``muonium''. Also, we reconsider the exotic muon decay
$\mu ^+\to e^+ +\overline \nu _e +\nu _{\mu}$ mediated by the dilepton gauge
boson. Using the experimental values of the Kobayashi-Maskawa (KM) matrix
and the unitarity relation of the KM matrix for the 3-family case,
we derive a lower bound on the singly-charged dilepton mass which is
numerically close to the presently most stringent one obtained
in Ref.\Carlson.

The dilepton interaction with leptons is given by \Framb\
\eqn\Lint{\eqalign{L_{\rm int}&=-{g_{3l}\over2\sqrt2}X_\mu^{++}l^T C
\gamma^\mu \gamma_5 l
-{g_{3l}\over2\sqrt2}X_\mu^{--}\overline l\gamma^\mu \gamma_5 C\overline l^T
\cr
&+{g_{3l}\over2\sqrt2}X_\mu^{+} l^T C
\gamma^\mu (1-\gamma_5)\nu_l
+{g_{3l}\over2\sqrt2}X_\mu^{-} \overline {\nu_l}
\gamma^\mu (1-\gamma_5) C\overline l^T,\cr}}
\noindent
where $l=e,\mu ,\tau $, and $C$ is the charge-conjugation matrix.
The gauge coupling constant $g_{3l}$ is
given approximately by $g_{3l}=1.19e$ for the SU(15) GUT model \Frama\
and by  $g_{3l}=g_2=2.07e$ for the
$SU(3)_L \times U(1)_X$ model \Framd,
where $e$ and $g_2$ are the electric charge and the $SU(2)_L$ gauge
coupling constant, respectively.
It is noted that the vector currents which couple to
doubly-charged dileptons
$X^{\pm \pm}$ vanish due to Fermi statistics.

\medskip
\noindent
(1) {\it The {\rm 1S} hyperfine splitting in the
muonium-antimuonium system.}

The hyperfine splitting between the spin-0 and spin-1 levels of the muonium
ground state has been measured to give a very precise result
\ref\Hyper{F. G. Mariam et al., Phys. Rev. Lett. {\bf 49}, 993 (1982);
E. Klempt et al., Phys. Rev. D{\bf 25}, 652 (1982).}%
\eqn\Exp{\Delta \nu ({\rm exp}) = 4 463 302.88 (0.16) \ {\rm kHz}
 \ \ \  (0.036 {\rm ppm}).}
On the other hand, the uncertainty in the theoretical prediction for
the hyperfine splitting of the muonium ground state $\Delta \nu ({\rm th})$
is rather large. Using
the most accurate value of $\alpha$ available at present which was
obtained from the theory and measurement of the electron anomalous magnetic
moment, Kinoshita gave
\ref\Kino{T. Kinoshita, Preprint No. CLNS 93/1219 (1993).}
\eqn\Theo{\Delta \nu ({\rm th:SM})
= 4 463 303.41 (0.66) (0.06) (0.17) \ {\rm kHz}}
\noindent where the first and second errors reflect the uncertainties in the
measurements of muon mass and $\alpha$. The last error in Eq.\Theo\ is the
theoretical uncertainty coming from radiative corrections and
radiative-recoil corrections \Kino. It is noted that the above
$\Delta \nu ({\rm th:SM})$ is the Standard-Model prediction and
includes also the weak interaction corrections.

When the doubly-charged dileptons exist, the mixing of muonium $M(\mu^+ e^-)$
and antimuonium $\overline M(\mu^- e^+)$ arises through the diagram
illustrated in
\fig\muo{The doubly-charged dilepton exchange diagram for
muonium-antimuonium transition. The arrows show the flow of lepton number.}.
The doubly-charged-dilepton-exchange diagram is effectively
described by the following Hamiltonian \Fujii
\eqn\Hint{H_{\rm eff}=A{[\overline \mu \gamma_\lambda (1-\gamma_5)
e][\overline \mu \gamma^\lambda (1+\gamma_5)e]+ H.c.}}
\noindent
where $A=-g_{3l}^2/(8M_{X^{\pm \pm}}^2)$ and
$M_{X^{\pm \pm}}$ is the doubly-charged dilepton mass. This form
is obtained from Eq.\Lint\ and with help of the Fierz transformation.
It should be noted that the above effective Hamiltonian
is in the $(V-A)\times(V+A)$ form, but not in the form of
$(V-A)\times(V-A)$ which was postulated in the work of Feinberg and Weinberg
\ref\Wein{G. Feinberg and S. Weinberg, Phys. Rev. {\bf 123}, 1439 (1961).
See also P. Herczeg and R. N. Mohapatra, Phys. Rev. Lett. {\bf 69}, 2475
(1992).}
, nor $(V+A)\times(V+A)$ which arises in the models with
doubly-charged Higgs bosons
\ref\Chang{D. Chang and W.-Y. Keung, Phys. Rev. Lett. {\bf 62},
2583 (1989);
 M. L. Swartz, Phys. Rev. D{\bf 40}, 1521 (1989).}
{}.
In consequence, we obtain a different value for the transition amplitude
$<\overline M\vert H_{\rm eff}\vert M>=\delta /2$ in the
triplet 1S hyperfine state from the one in the singlet state.
Specifically, we obtain
\eqn\Mix{\delta =\cases{-8A/\pi a^3,&for triplet state\cr
                         24A/\pi a^3,&for singlet state,\cr}}
\noindent
where $a$ is the Bohr radius of the muonium $(m_r\alpha )^{-1}$ with
$m_r^{-1}=m_{\mu}^{-1}+m_e^{-1}$, and $m_{\mu}$ and $m_e$ are muon and
electron masses, respectively.

Since there is a mixing between the muonium and antimuonium
states, the 1S hyperfine states of the muonium (or antimuonium) themselves
are not energy eigenstates any more. The mass matrix for
the triplet 1S state, for example, is written in the $M-\overline M$ space as
\eqn\Masss{{\cal M}_t =\left(\matrix{E_t&{\delta _t\over 2}\cr
                                     {\delta _t\over 2}&E_t\cr}\right),}
\noindent where the subscript {\it t} indicates the triplet state and
$\delta _t =-8A/\pi a^3$.
The eigenvalues of ${\cal M}_t$ and corresponding eigenstates are
\eqn\eigent{E_{t\pm} = E_t \pm {\delta _t\over 2},}
\eqn\state{\psi _{t\pm} ={1\over \sqrt 2}
(\psi _{M_t} \pm \psi _{\overline M_t}).}
Likewise, we obtain for the eigenvalues of the mass matrix for
the singlet 1S state and corresponding eigenstates,
\eqn\eigens{E_{s\pm} = E_s \pm {\delta _s\over 2},}
\eqn\state{\psi _{s\pm} ={1\over \sqrt 2}
(\psi _{M_s} \pm \psi _{\overline M_s})}
\noindent with $\delta _s =24A/\pi a^3$.
Thus, we find that in the models which
accomodate the dilepton gauge bosons,
the hyperfine splitting in the ``muonium'' ground state is predicted as
\eqn\Pred{\Delta \nu ({\rm th}) = \Delta \nu ({\rm th:SM})
+ \Delta \nu ({\rm DL}),}
\noindent where $\Delta \nu ({\rm th})=(E_{t\pm}-E_{s\pm})/ h $ and
$\Delta \nu ({\rm th:SM})=(E_t-E_s)/ h $, and $\Delta \nu ({\rm DL})$
is given by
\eqn\thDL{\Delta \nu ({\rm DL}) = {1\over {2h}}(\pm \delta _t
\pm \delta _s).}

Using the results Eqs.\Exp, \Theo\ and \Pred, we find that the difference
between the theory and measurement is
\eqn\diff{\Delta \nu ({\rm th})-\Delta \nu ({\rm exp})
= 0.53 + \Delta \nu ({\rm DL}) \pm 0.70 \ \ \ {\rm kHz},}
\noindent which implies that $\Delta \nu ({\rm DL})$ is bounded as
\eqn\bound{-1.23 < \Delta \nu ({\rm DL}) < 0.17 \ \ \ {\rm kHz}.}
\noindent This gives the limit
\eqn\bou{(\vert {\delta _t}\vert + \vert {\delta _s}\vert ) =
{{32\vert A\vert}\over {\pi a^3}} < 2 h \times 0.17 \ \ {\rm kHz},}
\noindent which is tranlated into the lower bound for the doubly-charged
dilepton mass
\eqn\boundb{{M_{X^{\pm \pm}}\over g_{3l}}>215 \rm GeV.}
\noindent When we use $g_{3l}=1.19e$ for the $SU(15)$ GUT model
($g_{3l}=2.07e$ for the $SU(3)_L \times U(1)_X$ model), we get
$M_{X^{\pm \pm}} > 77{\rm GeV} (135{\rm GeV})$.

Although our analysis could not give a better bound on the doubly-charged
dilepton mass than the one in Ref.\Framb,  the study of dilepton contributions
to
the ground state hyperfine splitting in the muonium-antimuonim system is
very important at least in two respects. Firstly, it should be noted that
when the dileptons
exist and there is a mixing between the muonium and anti-muonium states,
the 1S hyperfine states of the muonium (or antimuonium) themselves
are not energy eigenstates any more. Both the triplet and singlet 1S states
break up into further fine structures. Thus the ground states of the
muonium-antimuonim system are made up of four different energy eigenstates.
What we observe as a 1S hyperfine splitting in muonium is in fact the
energy difference between these energy eigenstates which are the
mixture of the muonium and anti-muonium states. Secondly, the muonium
hyperfine structure will provide one of the most stringest tests
of quantum electrodynamics (or Standard Model) in the near future.
The new experiment being undertaken at Los Alamos National Laboratory
is expected to improve the precision of $\Delta \nu ({\rm exp})$ by
a factor between 5 and 10
\ref\Hughes{V. W. Hughes, Z. Phys. C{\bf 56}, 35 (1992).}
. This experiment and further improved theoretical (QED)
predictions may serve not only as the high precision test of
QED but also as a probe for new physics beyond the Standard Model.

\medskip
\noindent
(2) {\it Exotic muon decay.}

The standard model forbids the muon decay to the ``wrong neutrino'',
\eqn\wrong{\mu ^+\to e^+ +\overline \nu _e +\nu _{\mu}}
which violates the individual lepton number conservation law but satisfies
a multiplicative lepton number conservation law. The existence of dileptons
and the interaction given in
Eq.\Lint\ make the above exotic muon decay mode possible.

Sometime ago Marciano and Sirlin
\ref\Marc{W. J. Marciano and A. Sirlin, Phys. Rev. D{\bf 35}, 1672 (1987).}
used the unitarity relation of the KM matrix for the 3-family
case and succeeded in providing the mass bounds for
the additional neutral gauge bosons.
The same idea was applied by Herczeg
\ref\Herc{P. Herczeg, Z. Phys. C{\bf 56}, 129 (1992).}
to provide a bound on the mass and couplings of non-standard Higgs bosons
which mediate the exotic muon decay
$\mu ^+\to e^+ +\overline \nu _e +\nu _{\mu}$. We will follow the same
prescription of
Refs.\Marc\Herc, and obtain the lower bound on the singly-charged dilepton
mass.

The first-row elements of KM matrix, $V_{uj}$ with $j=d, s, b$, have
been well determined experimentally by comparing hadronic $\beta$ decays
with muon decay rate. After the radiative corrections are taken into account,
they give
\ref\Marcb{W. J. Marciano, Annu. Rev. Nucl. Part. Sci. 1991.41: 469-509.}
\eqn\KM{ \vert V_{ud}\vert ^2 +\vert V_{us}\vert ^2 +
\vert V_{ub}\vert ^2 = 0.9991 \pm 0.0016.}
\noindent In the presence of the interaction Eq.\Lint, the l. h. s. of
the above relation is replaced with
$ \sum_{j=d,s,b}\vert V_{uj}\vert ^2 /(1+\Delta)$ \
where $\Delta=(g_{3l}M_W/g_2M_{X^{\pm}})^4$. Thus we obtain
\eqn\KMb{ \vert V_{ud}\vert ^2 +\vert V_{us}\vert ^2 +
\vert V_{ub}\vert ^2 = 0.9991 \pm 0.0016 + \Delta.}
The unitarity relation for the 3-family case then implies
\eqn\del{\Delta \leq 0.0029 \ \ \ \ \ (90\% {\rm C.L.}),}
\noindent which is translated into
\eqn\boundc{{M_{X^{\pm }}\over g_{3l}}>550 \rm GeV\ \ \ \ \
(90\% {\rm C.L.}) .}
\noindent If we use $g_{3l}=1.19e$ for the $SU(15)$ GUT model
($g_{3l}=2.07e$ for the $SU(3)_L \times U(1)_X$ model), we obtain
$M_{X^{\pm }} >200{\rm GeV} (350{\rm GeV})$.
This limit for the singly-charged dilepton
mass is numerically close
to the presently most stringent one obtained in Ref.\Carlson.

In conclusion, we have studied the doubly-charged dilepton gauge boson
contributions to the ground state hyperfine splitting in ``muonium".
When the dileptons
exist and there is a mixing between the muonium and anti-muonium states,
the 1S hyperfine states of the muonium (or antimuonium) themselves
are not energy eigenstates any more, and the triplet and singlet 1S states
break up into further fine structures. Next, we have obtained a lower
bound $(M_{X^{\pm }}/g_{3l})>550 \rm GeV$
for the mass of the singly-charged dilepton,
considering its contributions to the exotic muon decay
$\mu ^+\to e^+ +\overline \nu _e +\nu _{\mu}$ and
using the unitarity relation of
the KM matrix for the 3-family case.

\bigskip
\bigskip

We are grateful to Professor T. Kubota for suggesting the muonium-antimuonium
mixing. We are also grateful to Professor W. Marciano for introducing
the work of Ref.\Marc\ to us and for invaluable instructions. One of us (K.S)
would also like to thank Professor W. Marciano and Professor R. Pisarski
for the hospitality extended to him at Brookhaven National Laboratory
in the summer of 1993 when
the final stage of this work has been completed.

\listrefs
\listfigs
\bye